\documentclass[12pt,preprint]{aastex}
\shorttitle{Radiation from the Relativistic Jet}
\shortauthors{Stawarz \$ Ostrowski}
\usepackage{natbib}

\begin{document}

\title{Radiation from the Relativistic Jet:  \\
 a Role of the Shear Boundary Layer}

\author{\L . Stawarz and M. Ostrowski}
\affil{Obserwatorium Astronomiczne, Uniwersytet Jagiello\'{n}ski, \\
ul. Orla 171, 30-244 Krak\'{o}w, Poland}
\email{stawarz@oa.uj.edu.pl}

\begin{abstract}
Recent radio and optical large scale jets' observations suggest a two-component jet morphology, consisting of a fast central spine surrounded with a boundary layer with a velocity shear. We study radiation of electrons accelerated at such boundary layers as an option for standard approaches involving internal shocks in jets. The acceleration process in the boundary layer yields in a natural way a two component electron distribution: a power-law continuum with a bump at the energy, where energy gains equal radiation losses, followed by a cut-off. For such distributions we derive the observed spectra of synchrotron and inverse-Compton radiation, including comptonization of synchrotron and CMB photons. Under simple assumptions of energy equipartition between the relativistic particles and the magnetic field, the relativistic jet velocity at large scales and a turbulent character of the shear layer, the considered radiation can substantially contribute to the jet radiative output. In the considered conditions the synchrotron emission is characterized by a spectral index of the radio-to-optical continuum being approximately constant along the jet. A characteristic feature of the obtained broad-band synchrotron spectrum is an excess at X-ray frequencies, similar to the one observed in some objects by Chandra. As compared to the uniform jet models, the velocity shear across the radiating boundary region leads to decrease and frequency dependence of the observed jet-counterjet radio brightness asymmetry. We conclude that a careful investigation of the observational data looking for the derived effects can allow to evaluate the role of the boundary layer acceleration processes and/or impose constraints for the physical parameters of such layers in large scale jets.
\end{abstract}

\keywords{acceleration of particles---galaxies: jets---radiation mechanisms: nonthermal}

\section{Introduction} 

Jet radial stratification was proposed in order to interpret some radio and optical observations of  extragalactic jets. The existence of the jet boundary layer with a velocity shear was first suggested by the observations of jets in FR I sources. \citet{owe89} concluded that the jet morphology in M87 is dominated by the boundary layer more than by the shock structure. The constancy of the radio-to-optical spectral index along the jet and the similarity of radio and optical maps with emissivity peaking near the jet's surface strongly suggested the acceleration of radiating electrons {\it in situ} at the boundary region. Later polarimetry of M87 jet \citep{per99} confirmed its complex structure, which contains knots with a transverse magnetic field dominating optical emission, and a shear layer which is bright in the radio band with a strong polarization suggesting a highly ordered magnetic field parallel to the jet's axis. Jet-counterjet surface brightness asymmetries and magnetic field structures of the other FR I radio galaxies, 3C 31 and 3C 296 (\citealt{lai96} and \citealt{hac97}, respectively), were interpreted in terms of a model in which the jets consisted of a relativistic {\it spine} surrounded by a cylindrical layer with a velocity shear. A study of optical counterparts of the radio jets in BL Lac objects PKS 2201+044, PKS 0521-365, and 3C371 \citep{sca99} emphasized the problem of the reacceleration of electrons radiating within the optical band in the equipartition magnetic field, as the observed constant synchrotron spectral index along the jet could not be explained by the first-order Fermi shock acceleration. A map of the radio spectral index of another well known blazar, Mkn 501, revealed a boundary region of its jet with an unexpectedly flat spectrum \citep{edw00}. Once again, this suggested action of some reacceleration mechanism at the jet surface.

Some FR II radio sources also reveal a {\it spine - shear boundary layer} jet morphology. For instance, radio observations of 3C353 \citep{swa98} were interpreted in terms of a Doppler-hidden relativistic spine surrounded by the slower moving boundary layer. Polarimetry has shown no magnetic field component transverse to the jet axis in the boundary layer, in agreement with previous observations of radio linear polarization of quasars (e.g. \citealp{caw93}). The magnetic field being parallel to the flow velocity is expected in the presence of a strong velocity shear (e.g., \citealp{kah83}). VLBI observations of another FR II object 1055+018 \citep{att99} revealed a fragmentary but distinct boundary layer with the mean longitudinal magnetic field present where interaction with the surrounding matter takes place (i.e., where the jet bends). Optical and radio observations of the quasar 3C 273 were interpreted in terms of a two component jet model, in which a fast-moving proper jet was surrounded by the slow-moving radio {\it cocoon} \citep{bah95}. Smooth changes of the radio-to-optical spectral index along the jet and a lack of correlation between the optical flux and the spectral index \citep{jes01}, strongly suggest reacceleration of the radiating electrons taking place within the whole jet body. \citeauthor{jes01} also emphasized the discrepancy between the light-travel time along the jet in 3C 273 and the lifetime of electrons responsible for its synchrotron optical emission. This discrepancy could not be removed by relativistic beaming or the sub-equipartition magnetic field.         

A difference between FR I and FR II radio sources is ascribed to respective jet power and the importance of its interaction with the surrounding medium \citep{ghi98}. In general, FR I jets are thought to decelerate and spread out on the tens-of-kpc scale with their apparent magnetic field dominated by the component made perpendicular due to propagating shocks (e.g., \citealt{lai96,lai99}). On the contrary, jets in FR II sources seem to remain relativistic and well collimated even if they are far away from the nucleus. At the same time, their apparent parallel magnetic field configuration results from the interaction with the ambient medium, thereby creating the shear layer with an approximately constant (limited) thickness and dominating the observed polarization (\citealt{caw93,swa98}). The relativistic velocity on a large scale leading to apparent superluminal motions is also present in intermediate objects like M87 \citep{bir99}.

The broad band spectral properties of BL Lacs and FR I radio galaxies give us another argument in support of a jet stratification. By comparing the multiwavelength observations of these two classes of AGNs in a framework of a unification scheme, \citet{chi00} found evidence for a significant boundary layer emission, dominating the radiative jet output in FR Is. It is because FR I nuclei are over-luminous as compared to the predictions of the unification scheme, and because the ratios of radio to optical luminosities in BL Lacs are inconsistent with the ones observed in FR I's. The inferred velocities of such boundary regions are significantly slower as compared to the velocities of the central spines, but still relativistic, in order to explain anizotropic emission observed from the FR Is' cores.   

One should note, that our knowledge of physical conditions at the relativistic jet boundary is still insufficient. Several types of viscosity for the relativistic jet with a shear layer which confines the beam due to viscous interaction with the ambient medium were investigated by \citet{baa80}. Cosmic ray viscosity in a boundary layer with a velocity shear which influences the hydrodynamics of the jet and particle distribution along its surface was studied by \citet{ear88} and \citet{jok89}. Analogous radiation viscosity was investigated by \citet{ara92}. The issue of jet stability with respect to magnetohydrodynamic Kelvin-Helmholtz (KH) instabilities of a boundary layer in a vortex sheet approximation was studied by \citet{fer78,fer80,fer81,hae79,hae83} and \citet{bir84}, among others. The influence of the finite thickness shear layer was considered by \citet{fer82} and \citet{bir91}. More recently, \citet{han96,han98} investigated the growth of KH instabilities in two component relativistic and nonrelativistic jets. Such instabilities at the interface of a fluid beam and an external medium can influence the apparent jet morphology \citep{fer78,hae79,ben80} and generate turbulent MHD waves which accelerate particles \citep{fer79,eil82}. \citet{fer79} demonstrated that a small amplitude Alfv\'en turbulence developed by the KH instabilities in extended radio lobes can accelerate electrons to high energies via the wave-particle resonant interactions. \citet{eil82} showed that MHD turbulence generated by the surface instabilities and accelerating the electrons is restricted to a narrow turbulent edge of the jet. During the last few years, the numerical modeling of relativistic jets provided some additional information on the nature of such transition layers. 3D simulations revealed the presence of a shear layer with a high specific internal energy and highly turbulent, subsonic and thin cocoon surrounding a relativistic flow \citep{alo99}. Simulations of radio emission from such jets \citep{alo00}, show a radial structure in both intensity and polarization. One should note that the reason for the generation of the turbulent shear layer in those simulations was numerical viscosity, which only qualitatively mimics the real viscous processes. 

Till the present time no one performed a more thorough study of both acceleration processes acting at jet boundaries and the resulting observational effects. Our present paper is intended to study the possibility that the turbulent boundary layers of large scale jets can substantially contribute to the radiative jet output. The modifications of the standard models of jet radiation arise from particle acceleration acting continuously within the considered region and kinematic effects due to the velocity shear. In the present study, we neglect a possible separate and different energetic electron population present within the jet spine. We concentrate on the boundary layer radiation, in order to evaluate a role of the electron acceleration processes in this layer and to constrain hardly known physical parameters of this region. Below, in Section 2 we discuss an acceleration mechanism creating at the jet boundary a power-law particle distribution with a much harder component at its high energy end. The radiation of such electrons through the synchrotron, synchrotron self-Compton and external Compton processes, modified by the kinematic effects connected with the jet velocity shear, are derived in Section 3. Consequences of the considered acceleration mechanism in the context of multiwavelength large scale jets observations are discussed in Section 4. Finally, a few conclusions are presented in the last Section 5.

\section{Acceleration of cosmic ray electrons}

Particle acceleration in a velocity shear layer was proposed in the early eighties by Berezhko with collaborators and then gradually developed (c.f. a review by \citealt{ber90}). The action of such a mechanism at relativistic shear layers occurring at side boundaries of relativistic jets was discussed by \citet{ost90,ost98,ost00}. The mechanism was considered to provide high energy cosmic rays and to be an important factor influencing the dynamics of relativistic jets in extragalactic radio sources, thereby forming {\it cosmic rays cocoons}. \citet{ost01} proposed that such an acceleration mechanism could provide a cosmic ray proton population being a substantial pressure factor in FR II radio source lobes. Below we discuss the application of the model of \citet{ost00} to the electron acceleration in the large scale jets.

\subsection{Maximum electron energy}

Let us assume that a thickness of the transition layer between the considered tens-of-kiloparsecs scale jet and the surrounding medium, $D$, is comparable to the jet radius $R_j \approx 1 \, {\rm kpc}$ (c.f. \citealt{owe89,swa98}, see also \citealt{alo99}), and that the magnetic field ${\bf B}$ frozen into tenuous plasma at and outside the boundary, is parallel on average to the flow velocity, ${\bf U}$. The scattering on the magnetic field irregularities in a turbulent medium with a perpendicular ($\equiv$ radial) mean velocity gradient results in the acceleration of energetic particles injected into the boundary layer.

In our simple approach the efficiency of the acceleration process in the turbulent shear layer depends on the velocity structure of the layer and the particle mean free path $\lambda$. In the considered case of the mean magnetic field aligned along the jet in a highly turbulent boundary layer, a particle gyroradius, $r_g$, can be taken for $\lambda$, and a relatively inefficient cross-field radial diffusion provides a particle escape mechanism from the acceleration region. For extremely high energy particles with $r_{g} > D$, the jet boundary can be approximated as a surface of a discontinuous velocity change. Such a particle moving near the jet boundary interacts with the magnetic field perturbations and ocassionally crosses the discontinuity thereby increasing its energy on average. Numerical simulations for `typical' jet parameters \citep{ost98} reveal that protons can reach energies above $10^{19} \, {\rm eV}$ in this mechanism. For particles with lower energies, $r_g < D$, the acceleration within a finite thickness shear layer has to be considered. In this case, we deal with two different acceleration processes, the well-known second order Fermi acceleration in the turbulent medium and the acceleration due to {\it cosmic ray viscosity} in a velocity shear at the boundary. The cosmic ray viscosity in the context of particle acceleration was discussed by Berezhko in the early eighties (see \citealt{ber90} and references therein) and then by \citet{ear88} and \citet{jok89} in the case of nonrelativistic flows, and by \citet{ost00} for relativistic flow velocities. Below, we consider these mechanisms to provide relativistic electrons radiating through synchrotron and inverse Compton processes. Because of rapid energy radiation losses at high energies, the electrons are always expected to satisfy the required condition $r_{g} \ll D$. Then, the acceleration time scale in a highly turbulent shear layer can be estimated \citep{ost00} as
\begin{equation}
T_{acc} \sim \left({r_{g} \over c} \right) {c^{2} \over V^{2} + \left({r_{g} \over D} \right)^{2} U^{2}} , \eqnum{2.1}
\end{equation}
where $V$ is a characteristic velocity of the magnetic turbulence, $U$ is the jet velocity and we put $r_g$ for $\lambda$. The first term in the denominator represents the ordinary Fermi mechanism and the second one represents the viscous acceleration. The Fermi process dominates when the particle energy is small enough to satisfy $r_{g} < D (V / U)$.

We consider the jet velocity to be comparable to that of the light velocity, $c$, even on large scales. The velocity $V$ is comparable to the Alfv\'{e}n velocity, $V_{A}$, for a subsonic turbulence and possibly a few orders of magnitude smaller than $U$. MHD turbulent modes in a shear layer can be dominated by Alfv\'en waves, because the other types of waves are damped much more rapidly. In order to estimate the Alfv\'{e}n velocity, $V_{A} = B \, / \, \sqrt{4 \pi \rho}$, we assume that the mass density $\rho$ of the jet and its extended lobes is dominated by nonrelativistic protons, $\rho = m_{p} n$ \citep{sik00}. With the anticipated parameters for the boundary layer of the tens-of-kpc scale jet -- $B \approx 10^{-5} \, {\rm G}$ \ and the proton number density $n \approx 10^{-4} \, {\rm cm^{-3}}$ \ (c.f. \citealt{fer79}) -- one gets $V_A \approx 2 \cdot 10^{8} \, {\rm cm/s}$. Then, with the condition $r_{g} < D (V / U)$ satisfied and $V = V_A$, the characteristic acceleration time scale (2.1) for electrons (`$e$') reads as
\begin{equation}
T_{acc,e} \sim {r_{g} \over c} \left( {c \over V} \right)^{2} = 5 \cdot 10^3 \, \gamma \, B_{\mu G}^{-1} \, V_{8}^{-2} \quad [{\rm s}] , \eqnum{2.2}
\end{equation}
where $\gamma$ is the electron Lorentz factor, $V_8 \equiv V_A / 10^8 \, {\rm cm \, s^{-1}}$ and $B_{\mu G}$ is the magnetic field in micro-Gauss units. Instead of $n$, we will use $V_A$ to parametrize the acceleration process, because the mass density is not an observed quantity. One should note, that the equation (2.2) introduces an optimistic scenario for the acceleration process involving the large amplitude MHD turbulence at the scale of $r_g$, leading to short scattering free path $\lambda \sim r_g$. The time scale for particle escape from the acceleration region due to cross-field diffusion can be evaluated as :
\begin{equation}
T_{esc} \sim {D^2 \over \kappa_{\bot} } = 6 \cdot 10^{23} \, \gamma^{-1} \, \eta^{-1} \, B_{\mu G} \, D_{kpc}^2 \quad [{\rm s}] , \eqnum{2.3}
\end{equation}
where $\kappa_{\bot}$ is the cross-field diffusion coefficient, $\kappa_{\bot} = \eta \, r_g \, c / 3$, with $\eta \leq 1$ being a numerical scaling factor, and $D$ is the acceleration region (boundary layer) thickness ($D_{kpc} \equiv D / 1 \, {\rm kpc}$). Particle escape determines the electron energy spectrum, when the electrons satisfy the condition $T_{acc,e} \approx T_{esc}$ , i.e., when $r_g \approx (3 / \eta)^{1/2} D (V / c)$. For the considered boundary layer parameters and a strong turbulence condition $\eta \approx 1$, radiative cooling becomes important at much lower electron energies. We consider radiative losses due to the synchrotron and the inverse-Compton radiation (in the Thomson regime), to yield the loss time scale
\begin{equation}
T_{loss} \sim {6 \pi m c \over \sigma_{T} \gamma B^{2} (1+X)} = 8 \cdot 10^{20} \, \gamma^{-1} \, B_{\mu G}^{-2} \, (1+X)^{-1} \quad [{\rm s}] , \eqnum{2.4}
\end{equation}
where $X$ is a ratio of the photon field energy density to the magnetic field energy density. Therefore, the maximum electron energy in the acceleration process can be estimated by comparing the time scale (2.2) with the scale for radiative losses, (2.4). For a value of $X \leq 1$, possibly valid at the considered large distance from the active galactic nucleus (see next section), the maximum electron energy given by the condition $T_{acc,e} = T_{loss}$ is
\begin{equation}
\gamma_{eq} \approx 4 \cdot 10^{8} \, V_{8} \, B_{\mu G}^{-1 / 2} . \eqnum{2.5}
\end{equation}
In our model, a rapid radiative cooling of electrons with such high maximum energy is compensated by acceleration taking place along the jet within the whole boundary layer volume. For lower turbulence amplitude, the acceleration process can proceed slower to yield a lower value for $\gamma_{eq}$. One may note, that at $\gamma = \gamma_{eq}$ the ratio $T_{esc} / T_{acc} \approx 7 \cdot 10^2 D_{kpc}^2 B_{\mu G}^3$ ($\approx 7 \cdot 10^5$ for the considered jet parameters).

\subsection{Electron energy distribution}

A simple time dependent model of \citet{ost00} uses a regular acceleration term to represent continues second-order Fermi acceleration plus the not-considered here oblique shocks possibly formed at the jet side boundary. After switching on the injection of seed electrons with the low initial Lorentz factor $\gamma_0$, a spectral evolution of the distribution function $n_e(\gamma,t)$ for particles accelerated in the shear layer begins as a power-law with a growing energy cut-off. Then, either it forms a stable high energy cut-off when particle escape becomes substantial at some $\gamma < \gamma_{eq}$, or it evolves into a characteristic shape consisting of two different components: a power-law part at low energies, $n_e(\gamma) \propto \gamma^{-\sigma}$, finished with a hard component modeled here as a nearly monoenergetic peak at $\gamma = \gamma_{eq}$ due to the accelerated particles' pile-up caused by losses (Fig.~1). Normalization of the electron energy distribution depends on an unknown electron injection rate and is a free parameter of the model. In principle, the particle high energy component can grow to large values, limited only by the inefficient side way particle diffusion. However, when the energetic particle pressure becomes comparable to the ambient medium pressure, it can modify the transition layer structure. Then, some non-diffusive cosmic ray transport (turbulent diffusion, convective motions) or other non-linearities of the process can stabilize the average spectral distribution at the boundary on long time scales, or result in its locally  intermittent character. Below, because of the very nature of the acceleration process, we consider the stationary avarage electron spectrum formed within the boundary layer consisting of two basic components: a flat power-law component at lower energies and a high energy bump. Using the Haeviside function, $\Theta$, and the Dirac delta function, $\delta$, at energies above the injection energy, $\gamma > \gamma_0$, the considered electron distribution (cf. Fig.~1) can be expressed by an analytic formula
\begin{equation}
n_{e}( \gamma) = a \, \gamma^{- \sigma} \, \Theta( \gamma_{eq} - \gamma) + b \, \delta( \gamma - \gamma_{eq}) , \eqnum{2.6}
\end{equation}
where $a$ and $b$ are normalization parameters, and we put $\sigma = 2$. In the following discussion, we use numerical values for $a$ and $b$ providing equipartition between the magnetic field energy density $u_B \equiv B^2 / 8 \, \pi$ and the energy densities of both electron spectral components:
\begin{equation}
\int_{\gamma_0}^\infty (\gamma \, mc^2) \, a \, \gamma^{-2} \, \Theta( \gamma_{eq} - \gamma)  \, d \gamma = \int_{\gamma_0}^\infty  (\gamma \, mc^2) \, b \, \delta( \gamma - \gamma_{eq}) \, d \gamma = {1 \over 2} \, u_B , \eqnum{2.7}
\end{equation}
where $\gamma_{eq} > \gamma_0$. With assumed $B = 10^{-5} \, \rm{G}$ and maximum electron energy much higher than the injection energy, $\gamma_{eq} \sim 10^8 \gg \gamma_0 \sim 10 - 100$, the equipartitional normalization of the power-law electron distribution and the number density of the monoenergetic peak electrons are $a \approx 10^{-7} \, {\rm cm^{-3}}$ and $b \approx 10^{-14} \, {\rm cm^{-3}}$, respectively.

Let us mention, that an analogous model was discussed by \citet{sch84}, who considered similar pile-up mechanism for the relativistic electron distribution and studied evolution of the formed monoenergetic electron peak by solving stationary and time-dependent diffusion equation in the  momentum space. In his model, the first-order shock acceleration accompanied by the second-order turbulent Fermi acceleration creates the pile-up growing bump at the energy, where the radiative losses balance the energy gains. \citeauthor{sch84} assumed the mean free path of the electron being independent on its energy, contrary to the considered by us $\lambda \propto \gamma$. It was shown, that the stationary two-component electron spectrum (the power-law plus the monoenergetic peak) forms in a presence of efficient shock acceleration, when the escape time scale is much longer than the Fermi acceleration scale. After switching off the shock acceleration,  at the late phase of the evolution the high energy electron peak evolves into the ultrarelativistic Maxwell-Boltzmann distribution with the absolute height of the bump decreasing as $\exp ( - t / T_{esc} )$.

Below, we assume that the (quasi--) stationary averaged spectrum of electrons formed at the jet boundary layer can be represented by the distribution (2.6). As long as the diffusive particle escape to the sides is inefficient, the discussed acceleration process will create a high energy particle pile-up in a natural way (cf. \citealt{sta01}). The distributions produced within the jet boundary layer can vary in time and space reflecting the changes of the local conditions due to the jet interaction with the surrounding medium and action of non-linear effects in the acceleration process. However, the form of the averaged observed electron spectrum must consist of the two aforementioned components: the more or less exact power-law and the broadened by the fluctuating background high energy bump preceding the distribution cut-off. Without a detailed acceleration model available, depending also on the unknown velocity structure of the boundary layer, we use the simple form (2.6) to discuss the role of these two spectral components for a particular example of comparable energies stored in both components.

\section{Radiation of shear layer electrons}

We consider two radiation processes taking place at the relativistic jet boundary -- the synchrotron (SYN) emission of ultrarelativistic electrons with the energy spectrum (2.6), and their inverse Compton (IC) cooling. In the case of magnetic bremsstrahlung, we assume randomly orientated magnetic fields within the turbulent medium of the transition layer. A synchrotron self-absorption is neglected (i.e., the assumed optical depth for this process $\tau_{syn} \equiv \mu_{syn} l \ll 1$, where $\mu_{syn}$ is a synchrotrotron self-absorption coefficient and $l$ is an emitting region size). Therefore, in our calculations we limit the considered frequencies somewhat arbitrarily to the `safe' range $\nu \geq \nu_{abs} = 10^{10} \, {\rm Hz}$, where self-absorption effects are unlikely to occur in the considered large scale jets. Additionally, the present analytic IC scattering calculations are limited to the Thomson regime, assuming a step scattering cross section $\sigma = \sigma_T$ for $\gamma \, \epsilon_{i} < 1$ and 0 otherwise, where $\epsilon_{i}$ is the seed photon energy expressed in units of the electron rest energy, $mc^2$. A high energy spectral cut-off due to absorption of gamma rays via photon-photon pair creation on Cosmic Infrared Background (CIB) photons is expected to occur for distant sources (e.g., \citealt{ren01}). As we do not consider such processes, the obtained spectral distributions above TeV energies are the ones derived for close vicinity of the radiating jets and cannot be directly compared to observations. Considering the source of the target photons in a tens-of-kpc (or larger) scale jets, we neglect a  possibility of comptonization of the galactic narrow line radiation and IR dust emission \citep{cel01}, and we study the synchrotron self-Compton (SSC) radiation and the Compton scattering of external CMB photons (EC). With the assumed $B = 10^{-5} \, {\rm G}$, the magnetic field energy density at the jet boundary is $u_B = B^2 / 8 \pi \approx 4 \cdot 10^{-12} \, {\rm erg/cm^3}$. The energy density of the CMB radiation in a source frame moving with a bulk Lorentz factor $\Gamma$ (neglecting the cosmological redshift correction) is $u_{cmb} = u_{cmb}^{\ast} \Gamma^2 \approx 4 \cdot 10^{-13} \, \Gamma^2 \, {\rm erg / cm^3}$, where an asterisk denotes quantities in the observer's rest frame. In the jet rest frame, $u_{cmb}$ can be of the same order of magnitude as $u_B$, and both processes can be significant even for moderate $\Gamma$. One should note that in such a case, the estimate (2.5) for $\gamma_{eq}$ still holds. \citet{cel01}  considered also  comptonization of the blazar (`bl') emission by electrons in the boundary region of the kpc-scale conical jet, with a low average boundary layer Lorentz factor $\Gamma$. The energy density of such radiation as seen in the boundary layer local rest frame is $u_{bl} = L_{bl}' \, \Gamma_{pc}^2 \, / \, 4 \pi c \, z^2 \, \Gamma^2 \approx 3 \cdot 10^{-10} \, z_{kpc}^{-2} \, \Gamma^{-2} \, {\rm erg / cm^3}$, where $L_{bl}' \sim 10^{43} {\rm erg/s}$ is the intrinsic synchrotron blazar luminosity,  $\Gamma_{pc} \sim 10$ is the bulk Lorentz factor of the parsec-scale jet, and $z$ is a distance from the nucleus  ($z_{kpc} \equiv z / {\rm kpc}$). The blazar emission dominates over the magnetic field energy density at distances $z_{kpc} < 10 / \Gamma$, and over the CMB field at the scales $z_{kpc} < 30 / \Gamma^2$. Hence, for the studied large scale jets ($z \geq 10 \, {\rm kpc}$) and the relativistic flows ($\Gamma \geq 2$) the blazar emission is neglected. In order to derive analytic formulae for emission we are forced to make a simplifying approximation of the isotropic distribution of the synchrotron radiation in the source local rest frame. In reality, the shear layer synchrotron radiation distribution is anizotropic due to kinematic effects of the nonuniform relativistic flow. A numerical modeling of such anizotropic radiation field will be presented in a forthcoming paper.

\subsection{Synchrotron radiation}

The synchrotron emissivity of an isotropic ultrarelativistic electron distribution $n_e (\gamma)$ averaged over a randomly orientated magnetic field, $B$, is
\begin{equation}
j_{syn}(\nu) = { \sqrt{3} e^{3} B \over mc^{2}} \int R \left( {\nu \over c_{1} \gamma^{2}} \right) \, {n_{e}(\gamma) \over 4 \pi} \, d\gamma , \eqnum{3.1}
\end{equation}
where $c_1 = 3 e B \, / \,  4 \pi m c$ and $R(x)$ is given by a combination of the modified second order Bessel functions \citep{cru86} as
\begin{equation}
R(x) = {x^2 \over 2} \, K_{4/3} \left({x \over 2} \right) \, K_{1/3} \left({x \over 2} \right) - 0.3 \, {x^3 \over 2} \left[ K_{4/3}^2 \left({x \over 2} \right) - K_{1/3}^2 \left({x \over 2} \right) \right] . \eqnum{3.2}
\end{equation}
Upon integrating the equation (3.1) for the electron distribution (2.6) and expanding the function $R(x_{eq})$ for $x_{eq} \equiv \nu / c_1 \, \gamma^2_{eq} \ll 1$, $R(x_{eq}) \approx 1.8 \cdot x_{eq}^{1/3}$, one can obtain simple analytic expression for the synchrotron emissivity. For $x_{eq} \gg 1$, the function (3.2) decreases exponentially, $R(x_{eq}) \propto \exp(-x_{eq})$, and thus we neglect emission at frequencies $\nu \gg c_1 \gamma^2_{eq}$. For parameters $a_{-7} \equiv a / (10^{-7} \, {\rm cm^{-3}})$ and $b_{-14} \equiv b / (10^{-14} \, {\rm cm^{-3}})$, the synchrotron emissivity is
\begin{equation}
j_{syn}(\nu)= \left[ f_1 \, \nu^{-1/2} + f_2 \, \nu^{1/3} \right] \, \Theta( \nu_{syn,eq} - \nu) , \eqnum{3.3}
\end{equation}
where
\begin{displaymath}
f_1 = 3 \cdot 10^{-36} \, B^{3/2}_{\mu G} \, a_{-7} \quad [{\rm cgs}] ,
\end{displaymath}
\begin{displaymath}
f_2 = 2 \cdot 10^{-43} \, \gamma_{eq}^{-2/3} \, B^{2/3}_{\mu G} \, b_{-14} \quad [{\rm cgs}] ,
\end{displaymath}
\begin{displaymath}
\nu_{syn,eq} = 1.3 \, B_{\mu G} \, \gamma_{eq}^{2} \quad [{\rm Hz}] .
\end{displaymath}
For illustration, let us compare the synchrotron radiation spectra of the particle distributions in the acceleration model of \citet{ost00} in successive times from the switching on of the acceleration process (Fig.~1). From the beginning of injection, the accelerated electrons form a power-law spectrum with an energy cut-off which grows with time. Then the resulting emission has (for $\sigma = 2$) a power-law form $ \propto \nu^{-1/2}$ with a growing cut-off frequency. When the maximum electron energy approaches $\gamma_{eq}$, a particle density peak starts to form and grow at $\gamma \approx \gamma_{eq}$. Then, an additional radiation component $\propto \nu^{1/3}$ appears and eventually starts to dominate the spectrum at highest synchrotron frequencies, reaching the maximum at $ \nu_{syn,eq} $ ($\sim 10^{17} \, {\rm Hz}$ for $\gamma_{eq} \approx 10^{8}$ and $B_{\mu G} \approx 10$).

\subsection{ Synchrotron Self-Compton radiation}

The inverse-Compton photon emissivity in the source frame, $\dot{n}_{ic} (\epsilon, \Omega)$, is given by
\begin{equation}
\dot{n}_{ic}(\epsilon,\Omega) = c \int d\epsilon_i \oint d\Omega_i \int d\gamma \oint d\Omega_e \, (1-\beta \, \cos \psi) \, \sigma \, n_i(\epsilon_i,\Omega_i) \, n_e(\gamma,\Omega_e) , \eqnum{3.4}
\end{equation}
where $\epsilon$ is the photon energy in the electron rest mass units, $\epsilon = h \nu / mc^2$, \ $n_i (\epsilon_i, \Omega_i)$ is the seed photon number density (we put the indices $i = syn$, $ic = ssc$ for SSC emission, and $i=cmb$, $ic = ec$ for EC emission), $n_e (\epsilon_e, \Omega_e)$ is the electron energy distribution, and $\psi$ is an angle between the electron and the incident photon directions (e.g., \citealt{der95}). All quantities are given in the source rest frame, which in our case, is a part of the boundary region -  a cylindrical layer moving with a constant Lorentz factor $\Gamma$ (see section 3.4). The rest-frame emissivity in cgs units can be found next as $j_{ic}(\nu,\Omega) = h \, \epsilon \, \dot{n}_{ic} (\epsilon, \Omega)$. 

Computations of the SSC emission are performed in the Thomson regime with a scattering cross section independent of the seed photon energy, and with the scattered photons beamed along the electron direction. The synchrotron radiation and the electron distribution are assumed to be isotropic in the source frame, $n_e (\gamma, \Omega_e) = n_e (\gamma) / 4 \pi$ and $n_{syn} (\epsilon_{syn}, \Omega_{syn}) = n_{syn} (\epsilon_{syn}) / 4 \pi$. Thus, the average energy of the comptonized synchrotron photon is $\epsilon = {4 \over 3} \gamma^2 \epsilon_{syn}$, and in equation (3.4) one can put $\sigma = \sigma_T \, \delta(\Omega - \Omega_e) \, \delta\left(\epsilon - {4 \over 3} \gamma^2 \epsilon_{syn} \right)$ to obtain 
\begin{equation}
\dot{n}_{ssc}(\epsilon,\Omega) = {c \sigma_T \over 4 \pi} \int d\epsilon_{syn} \int d\gamma \, n_{syn}(\epsilon_{syn}) \, n_e(\gamma) \, \delta\left(\epsilon - {4 \over 3} \gamma^2 \epsilon_{syn} \right) . \eqnum{3.5}
\end{equation}
For the electron energy distribution given by the equation (2.6) and the optically thin synchrotron intensity, $I_{syn}(\nu_{syn}) \equiv j_{syn}(\nu_{syn}) \, l = {c h \over 4 \pi} \, \epsilon_{syn} \, n_{syn}(\epsilon_{syn})$ (where $l$ is an emitting region linear size), putting $\ln \left(\nu_{max}/ \nu_{abs}\right) \sim 10$, where $\nu_{max} = \min ( \nu_{syn,eq} \, , \, mc^2 / h \gamma_{eq} ) = m c^2 / h \gamma_{eq} $, and neglecting absorption due to photon-photon pair production, one can obtain the isotropic SSC emissivity in the Thomson regime
\begin{equation}
j_{ssc}(\nu) = \left[ f_3 \, \nu^{-1/2} + f_4 \, \nu^{1/3} \right] \, \Theta ( \nu_{ssc,eq} - \nu) , \eqnum{3.6}
\end{equation}
where
\begin{displaymath} 
f_3 = 10^{-44} \, B_{\mu G}^{3/2} \, l_{kpc} \left( 3 \, a^2_{-7} + 10^{-7} \, \gamma_{eq} \, a_{-7} \, b_{-14} \right) \quad [{\rm cgs}] ,
\end{displaymath}
\begin{displaymath} 
f_4 = 5 \cdot 10^{-58} \, B_{\mu G}^{2/3} \, \gamma_{eq}^{-4/3} \, l_{kpc} \, b^{2}_{-14} \quad [{\rm cgs}] ,
\end{displaymath}
\begin{displaymath} 
\nu_{ssc,eq} = {4 \over 3} \, \nu_{max} \, \gamma_{eq}^2 = 1.6 \cdot 10^{20} \, \gamma_{eq} \quad [{\rm Hz}] .
\end{displaymath}
Note, that the high energy electron bump contributes to both components of the SSC emission (i.e., also to the one $\propto \nu^{-1/2}$).

\subsection{External Compton radiation}

Cosmic microwave background (CMB) intensity, $I_{cmb}(\nu)$, has a blackbody form with a temperature $T^{\ast}=2.73 \, {\rm K}$. In a source frame moving with the relativistic Lorentz factor $\Gamma$, the CMB radiation is strongly anizotropic, providing an external radiation flux mainly from the direction opposite to the jet direction. Hence, for ultrarelativistic electrons, one can put $(1-\beta \, \cos \psi) = (1 + \mu_e)$, where $\Omega_e = (\cos^{-1} \mu_e, \, \phi_e)$. Then, the scattered photon energy is $\epsilon = \gamma^2 \, \epsilon_{cmb} \, (1 + \mu_e)$. Therefore, in the Thomson regime $\sigma = \sigma_T \, \delta(\Omega - \Omega_e) \, \delta\left[\epsilon - \gamma^2 \epsilon_{cmb} \, (1 + \mu_e) \right]$, and for the isotropic electron distribution, the equation (3.4) reads as   
\begin{equation}
\dot{n}_{ec}(\epsilon,\Omega) = {c \sigma_T \over 4 \pi} \,(1 + \mu) \int d\epsilon_{cmb} \int d\gamma \, n_{cmb}(\epsilon_{cmb}) \, n_e(\gamma) \, \delta\left[\epsilon - \gamma^2 \epsilon_{cmb} (1 + \mu) \right] . \eqnum{3.7} 
\end{equation}
Considering the power-law component of the electron energy distribution, it is convenient to model the CMB field as a monochromatic radiation with photon energies $\epsilon_{cmb} = \Gamma <\epsilon_{cmb}^{\ast}> \equiv 2.7 \cdot \Gamma \, k \, T^{\ast} / m c^2$, and a number density $n_{cmb} = \Gamma \, u_{cmb}^{\ast} / (<\epsilon_{cmb}^{\ast}> mc^2)$. For the monoenergetic electron peak component, one should use the exact blackbody spectrum transformed to the source frame. Hence, the emissivity of the comptonized CMB radiation, $j_{ec}(\nu,\Omega)$, is
\begin{equation}
j_{ec}(\nu,\Omega) = \left[ f_5(\mu) \, \nu^{-1/2} + f_6(\mu) \, {\nu^3 \over \exp[f_7(\mu) \, \nu]-1} \right] \, \Theta(\nu_{ec,eq} - \nu) , \eqnum{3.8}
\end{equation}
where
\begin{displaymath} 
f_5(\mu) = 8 \cdot 10^{-41} \, a_{-7} \, [\Gamma \, (1+\mu)]^{3/2} \quad [{\rm cgs}] ,
\end{displaymath}
\begin{displaymath} 
f_6(\mu) = 10^{-85} \, \gamma_{eq}^{-6} \, b_{-14} \, (1+\mu)^{-2} \quad [{\rm cgs}] , 
\end{displaymath}
\begin{displaymath} 
f_7(\mu) = 3 \cdot 10^{-11} \, \Gamma \, (1+\mu)^{-1} \, \gamma_{eq}^{-2} \quad [{\rm Hz^{-1}}] ,
\end{displaymath}
\begin{displaymath} 
\nu_{ec,eq}(\mu) = 2 \cdot 10^{11} \, \Gamma \, (1+ \mu) \, \gamma_{eq}^2 \quad [{\rm Hz}] . 
\end{displaymath}
The high energy bump radiation with frequencies $\nu < f_7^{-1}(\mu)$ (i.e., anizotropic Rayleigh-Jeans part of the blackbody CMB spectrum comptonized by the monoenergetic electrons) has a power-law form $\propto \nu^2$.

\subsection{Kinematic effects}

Let us briefly discuss the main implications of the kinematic effects on the jet radiative output. First of all, the shear layer consists of sub-layers moving with different velocities and thus, generating different beaming patterns. In order to specify the jet velocity structure, we assume a uniform flow Lorentz factor $\Gamma_j$ inside the jet spine, i.e. for $0 < r < R_j$ where $r$ is a distance from the jet axis, and a radial profile $\Gamma = \Gamma(r)$ within the cylindrical transition region $R_{j} < r < R_{c}$, where the flow velocity decreases and becomes zero ($\equiv$ the ambient medium velocity) at the radius $R_{c}$. Let us consider the simplest case of the linear dependence 
\begin{equation}
\Gamma(r) = 1 + (\Gamma_j-1) {R_c - r \over R_c - R_j} . \eqnum{3.9}
\end{equation}
for $R_j \le r \le R_c$. In each considered sub-layer, in the `local source frame' the synchrotron, SSC and EC emissivities are approximately given by formulas (3.3), (3.6), and (3.8), respectively. Assuming that the radiating electrons are distributed uniformly (as seen in the local rest frames) inside the whole boundary region, one can find the observed emissivity of the sub-layer at the distance $r$ from the jet axis as
\begin{equation}
j^{\ast}_i (\nu^{\ast}, \theta^{\ast}, r) = \delta^2(r,\theta^{\ast}) \, j_i\left( \nu = {\nu^{\ast} \over \delta(r,\theta^{\ast} )}, \, \mu = { \mu^{\ast} - \beta(r) \over 1 - \beta(r) \, \mu^{\ast}} \right) , \eqnum{3.10}
\end{equation} 
where the index $i = syn, ssc$ or $ec$, and $\delta(r, \theta^{\ast})$ is a Doppler factor for a sub-layer moving with the Lorentz factor $\Gamma(r) = \left[ 1 - \beta^{2}(r) \right]^{-1/2}$ at an angle $\theta^{\ast} \equiv \cos^{-1} \mu^{\ast}$ with respect to the line of sight, $ \delta(r,\theta^{\ast}) = 1 \, / \, \Gamma(r) \left[1 - \beta(r) \, \mu^{\ast}\right]$. Next, the observed flux density from the boundary layer volume $V^{\ast}$ can be found as $S^{\ast} (\nu^{\ast}, \theta^{\ast}) = d^{-2} \int dV^{\ast} \, j^{\ast}(\nu^{\ast}, \theta^{\ast}, r)$, where $d$ is the distance to the observer. Noting that $\Gamma(r) [1+\mu] \equiv \delta [1+\mu^{\ast}]/[1+\beta(r)]$, and neglecting the slowly varying factor $[1 + \mu^{\ast}] / [1+\beta(r)]$ (c.f. \citealt{der95}) from the above equations one obtains the flow modified beaming patterns for the synchrotron, SSC, and EC radiation, as 
\begin{equation}
S^{\ast}_{syn} (\nu^{\ast}, \theta^{\ast}) \propto (\nu^{\ast})^{- \alpha} \int_{R_j}^{R_c} dr \, r \, \delta^{2+\alpha}(r, \theta^{\ast}) , \eqnum{3.11a}
\end{equation}
\begin{equation}
S^{\ast}_{ssc} (\nu^{\ast}, \theta^{\ast}) \propto (\nu^{\ast})^{- \alpha} \int_{R_j}^{R_c} dr \, r \, \delta^{2+\alpha}(r, \theta^{\ast}) \, / \, \sin \theta^{\ast} , \eqnum{3.11b}
\end{equation}
\begin{equation}
S^{\ast}_{ec} (\nu^{\ast}, \theta^{\ast}) \propto (\nu^{\ast})^{- \alpha} \int_{R_j}^{R_c} dr \, r \, \delta^{3+2 \alpha}(r, \theta^{\ast}) , \eqnum{3.11c}
\end{equation}
where $\alpha$ is the spectral index of the respective power-law emission, and $\alpha = -2$ for the Rayleigh-Jeans part of the comptonized CMB spectrum. In the case of SSC emissivity, we assumed isotropic distribution of the seed photons in the source rest frame, and we put the effective emitting region linear size scaling roughly like $l \propto D \, / \, \sin \theta^{\ast}$. Integration in (3.11) can be performed numerically for the given values of $\Gamma_j$, $R_j$, $R_c$, $\theta^{\ast}$, and for the assumed radial velocity profile.

\subsection{Composite spectra}

Our model of high energy radiation generated within the jet boundary region has a few free parameters. They include normalization of the power-law electron distribution, $a$, a height of the monoenergetic electron peak, $b$, the magnetic field induction, $B$, the Alfv\'{e}n velocity in the shear layer, $V_A$ (or the mass density of the plasma within the boundary region, $\rho$), and the injection electron energy, $\gamma_0$. In general, the electron power-law spectral index, $\sigma$, can also be treated as a free parameter represented in this paper by the model value $\sigma = 2$. Here, $V_A$ (or $\rho$) and $B$ limit a value of the maximum electron energy, $\gamma_{eq}$. Relative normalizations of radiation spectral components -- synchrotron, SSC and EC emission -- depend on the above parameters and hence the observed composite boundary layer spectral energy distribution varies with the selected values of $a$, $b$, $B$, $\gamma_{eq}$, $\gamma_0$ together with the geometrical factors $\Gamma_j$, $R_j$, $R_c$ and $\theta^{\ast}$.

For a `typical' tens-of-kpc scale jet the magnetic field and the Alfv\'{e}n speed are, say, $B \approx 10^{-5} \, {\rm G}$ and $V_A \sim 10^8 \, {\rm cm /s}$, and the maximum electron energy can be as high as $\gamma_{eq} \sim 10^8$ (Eq.~2.5). We assume, that in the boundary region $\gamma_0 \ll \gamma_{eq}$ and that the electrons are uniformly distributed within the whole radiating volume (as seen in the local layer rest frame). The magnetic field energy density can provide estimates for normalizations of both electron spectral components, if we require the equipartition: $a \approx 10^{-7} \, {\rm cm^{-3}}$ and $b \approx 10^{-14} \, {\rm cm^{-3}}$ (Eq.~2.7). The thickness of the large scale jet boundary layer and the jet radius are assumed to be of the order of a kiloparsec, $D \approx R_j \approx 1 \, {\rm kpc}$, and hence $R_c = 2 \, R_j$. Below, for illustration, we consider a part of the jet with the observed length $\Delta z \approx 1 \, {\rm kpc}$. We also assume, that the jet spine is highly relativistic even at large scales, with that the bulk Lorentz factor $\Gamma_j \approx 10$ (c.f. \citealt{ghi01}). The jet with the described parameters is used to illustrate possible role of kinematic effects at the observed brightness distribution along the jet.

The observed spectral energy distribution (SED) of the emission generated at the considered part of the boundary region, $\nu^{\ast} S^{\ast}(\nu^{\ast}, \theta^{\ast})$, is shown on figure 2 for four different viewing angles. We put the source distance from the observer $d \approx 10^4 \, {\rm kpc}$. In order to estimate the observed luminosity of the boundary layer emission, one has to integrate over the intrinsic luminosities of the sub-layers with volumes $dV^{\ast}\equiv 2 \pi \, \Delta z \, r \, dr$ and the Doppler factors $\delta (r, \theta^{\ast})$. For example, the synchrotron luminosity of the boundary layer electrons is
\begin{equation}
L_{syn}^{\ast}(\theta^{\ast}) = \int \delta^3 (r, \theta^{\ast}) \, dL_{syn} \equiv \int\!\!\!\int \delta^3 (r, \theta^{\ast}) \, \dot{\gamma}_{loss} \, mc^2 \, n_e(\gamma) \,d \gamma \, dV^{\ast} , \eqnum{3.12}
\end{equation} 
where $\dot{\gamma}_{loss} = \gamma / T_{loss}$ denotes the synchrotron energy losses. In our illustrative example, synchrotron luminosity of the high energy bump electrons observed at $\theta^{\ast} = 45^0$ (i.e. the observed flux multiplied by the $4 \pi d^2$) can be as high as $10^{42} \, {\rm erg/s}$. 

\section{Multiwavelength large scale jet boundary emission}

In this section we compare the model of a radiative boundary layer with the large scale jet observations. We do not consider the possible contribution from the separate population of relativistic electrons accelerated within the knots due to the first order shock acceleration. Our considerations are intended to describe only the main spectral characteristics of the relativistic electrons accelerated within the turbulent boundary region.

\subsection{Radio-to-optical continuum}

For the anticipated large scale jet parameters, assuming equipartition between the magnetic field and the relativistic electrons accelerated within the shear layer, one will find the power-law electron component dominating the radio-to-optical jet boundary emission, as illustrated on figure 2. As a result of a continuous acceleration taking place within the whole boundary layer, the radio-to-optical spectral index is constant along the jet. Of course, changes of the physical conditions with the distance from the nucleus (e.g., changes of the magnetic field, turbulent conditions, etc.) can result in smooth changes of the slope of the synchrotron continuum, like in the case of M 87 or 3C 273 (\citealt{mei96, jes01}, respectively). In our model the acceleration process creates the high energy electron spectral bump and the optical emission joins smoothly the X-ray bump emission. Such spectral character is different from the case usually considered in the literature of the power-law continuum ended by the cut-off at optical or X-ray frequencies, followed by the inverse-Compton component.

The important manifestation of the boundary layer radio-to-optical emission is a decrease of the jet-counterjet brightness asymmetry as compared to the uniform jet, suggested previously by \citet{kom90}. This effect arise from the fact, that in a presence of the velocity shear, for a given viewing angle $\theta^{\ast} > 1 / \Gamma_j$ the observed synchrotron emissivity is maximized for $\Gamma(r) = 1 / \sin \theta^{\ast} $ (i.e., for the maximum value of the Doppler factor $\delta (r, \theta^{\ast})$), while the uniform jet radiation, if present, is Doppler-hidden. Figure 3 illustrates the boundary layer jet-counterjet brightness asymmetry,
\begin{equation}
{ S^{\ast}_{syn} (\theta^{\ast}) \over S^{\ast}_{syn} (\pi - \theta^{\ast}) } = { \int_{R_j}^{R_c} dr \, r \, \delta^{2+\alpha}(r, \theta^{\ast}) \over \int_{R_j}^{R_c} dr \, r \, \delta^{2+\alpha}(r, \pi - \theta^{\ast}) } , \eqnum{4.1}
\end{equation} 
for the spectral index $\alpha = 0.5$ corresponding to the radio emission of our power-law electron component and the Lorentz factor profile (3.9) with $R_c = 2 \, R_j$ and $\Gamma_j = 10$ or $3$. For comparison, respective flux asymmetries for the uniform jet models are also plotted. Due to the considered effect, for $\Gamma_j = 10$ and the moderate inclinations, the observed jet asymmetry can be reduced by more than one order of magnitude in comparison to the uniform jets. Hence, the middly-relativistic velocities inferred from the observed jet asymmetries of the tens-of-kpc scale jets can correspond to the slower boundary layer and not necessarily to the fast spine, which can still be highly relativistic at the observed distances. 

\subsection{X-ray bump emission}

In the model presented here, the X-ray radiation of large scale jets is dominated by the high energy electron bump spectral component. As illustrated on figure 2, the observed synchrotron X-ray flux can be above the extrapolated radio-to-optical continuum. The spectrum inclination at X-ray frequencies can be different from the power-law observed at lower frequencies, and the difference depends also on a viewing angle. Figure 4 illustrates this dependence for the effective X-ray spectral index $\alpha_{x,eff}(\theta^{\ast})$ computed between $h\, \nu^{\ast}_1 = 1 \, {\rm keV}$ and $h\, \nu^{\ast}_2 = 5 \, {\rm keV}$ as
\begin{equation}
\alpha_{X, eff}(\theta^{\ast}) = { \log \left[ S^{\ast}_X (\nu^{\ast}_1, \theta^{\ast}) \, / \, S^{\ast}_X (\nu^{\ast}_2, \theta^{\ast}) \right] \over \log \left[ \nu^{\ast}_2 \, / \, \nu^{\ast}_1 \right] } . \eqnum{4.2}
\end{equation}
where $ S^{\ast}_X (\nu^{\ast}, \theta^{\ast}) \propto \int r \, dr \, \delta^2(r,\theta^{\ast}) \, R (x_{eq}^{\ast})$, a function $R(x_{eq}^{\ast})$ is given by equation 3.2, and $x_{eq}^{\ast} = \nu^{\ast} / c_1 \gamma^2_{eq} \delta(r, \theta^{\ast})$. The difference between $\alpha_{X,eff}$ for different jet Lorentz factors and different viewing angles results from the spectral character of the high energy electron bump emission, which is strongly concentrated around the maximum synchrotron frequency $\nu_{syc, eq}$. Transformation to the observer's rest frame, which depends on the flow Lorentz factor and on the inclination $\theta^{\ast}$, results in the fact, that for the fixed observed frequency $\nu^{\ast}$ in this spectral range ($\nu^\ast \sim \nu_{syn, eq}$ )different parts of the X-ray continuum can contribute to the radiative output. For the higher Doppler beaming factors (i.e., higher $\Gamma_j$ or smaller $\theta^{\ast}$) the effective X-ray spectral index of the bump electrons with $\gamma_{eq} \sim 10^8$ approach the asymptotic value $-1/3$. In the particular case considered by us the counterjet has a steeper X-ray continuum, as compared to the jet spectrum. Moreover, for large viewing angles, X-ray emission of the jet softens significantly, opposite to the counterjet emission. Note, that such a behavior is opposite to what we expect in the case of IC models involving the low-energy tail of the electron distribution, in which the X-ray spectral index of the jet and the counterjet should be the same, and approximately constant with the distance from the nucleus and with the viewing angle. The boundary X-ray emission at the large scales can therefore vary among different types of the jets, reflecting different physical conditions at their edges and different kinematic effects involved. Thus, it is interesting to note that the large scale X-ray emission observed from several radio loud AGNs exhibit the different spectral characteristics. 

{\it Chandra} X-ray Observatory has detected significant emission from the large scale jets in quasars (e.g. PKS 0637-752: \citealt{cha00}), FR II sources (e.g. Pictor A: \citealt{wil01}), and FR I radio galaxies (e.g. M 87: \citealt{mar01b}). In the majority of cases, thermal bramstrahlung can be excluded, and the non-thermal models seem to provide the only possible explanation for the X-ray emission \citep{har01}. However, the spectral index of the X-ray non-thermal continuum is different for different sources. Also the ratio of the X-ray flux to the optical flux changes significantly in the Chandra jets sample. This suggest, that different radiative processes can be at work. The models of the non-thermal large scale jet X-ray emission usually involve the synchrotron emission of high energy electrons, with Lorentz factor $\gamma \sim 10^7$, or the inverse Compton scattering of CMB photons by much less energetic electrons ($\gamma \sim 10^2$) in a relativistic jet. Both classes of models are subject to difficulties in explaining the variety of Chandra observations. The synchrotron emission from one population of the electrons accelerated in a relativistic shock is excluded if the X-ray flux is above the extrapolated radio-to-optical continuum. Also, the continuous along the jet character of the large scale X-ray emission is inconsistent with short lifetimes of the electrons radiating X-rays in the equipartition magnetic fields, unless we consider continuous acceleration within the whole jet body. On the other hand, the `beamed' external Compton (EC({\it beam})) models require highly relativistic jet flows on the tens-of-kpc scales, and extrapolation of the single power-law electron energy distribution to low energies, while, in most of the cases, there is no direct evidence supporting these assumptions. As discussed in detail by \citet{hak01} such models can be excluded if the X-ray spectrum is much steeper than the one observed at radio frequencies. One should also note, that the long life-time of the electrons scattering CMB photons is in contrast with the knotty morphology of the X-ray jets, as well as with the off-sets between the radio and the X-ray peaks reported in several brightest knots. 

The EC({\it beam}) model with significant Doppler beaming seems to work well in the case of the quasar PKS 0637-752 \citep{tav00}, where VLBI observations suggest the parsec-scale flow Lorentz factor $\Gamma_j > 18$ and the inclination angle $\theta^{\ast} < 6.4^0$. For other quasars the situation is more complicated. For instance, observations of the jet in 3C 273 \citep{mar01a,sam01} are insufficient to choose between the synchrotron or the EC origin of the X-ray knots' emission, and the existence of an additional electron population radiating in this range is not excluded. One should also note recently published HST observations \citep{jes02} showing smooth transition from optical through UV to X-ray frequencies, as expected in our model. The X-ray luminosities of the known quasars' jets are in the range $10^{43} - 10^{45} \, {\rm erg/s}$, and their X-ray spectra are very flat, i.e. $\alpha_X \sim 0.23$ for 3C 207, \citep{bru01}, $\sim 0.5$ for PKS 1127, \citep{sie02}, or $\sim 0.8$ for 3C 273 and PKS 0637. In our model, the observed luminosity of the X-ray emission generated at the boundary layer can reach the observed values for reasonable jet parameters (c.f. discussion in section 3.5), and the low effective spectral index is consistent with small jet inclinations (c.f. Fig.~4). The significant X-ray emission is also observed from the large scale jets in radio galaxies. The lower X-ray luminosities of such jets, lying in the range $10^{39} - 10^{42} \, {\rm erg/s}$, and their relatively steep X-ray spectra, $\alpha_X \sim 1.0 - 1.5$ (\citealt{kra00} for Cen A, \citealt{hac01} for 3C 66B, \citealt{wor01} for B2 0206 and B2 0755), correspond to the large inclination angles on figure 4 and/or lower jet Lorentz factors. In most of jets in the radio galaxies, the EC({\it beam}) models are excluded, as they require too high beaming factors, the magnetic fields much below the equipartition values and too steep low-energy electron spectra. On the other hand, the synchrotron radiation of the shock accelerated electrons cannot explain the strong diffusive emission extending few kpc from the nucleus in, e.g., Cen A and 3C 66B. 

We expect, that the detailed investigation of the knots emission can allow verification of our two-component electron spectrum model, if the knots form by the shock compression of the energetic electrons {\it previously} accelerated at the boundary layer upstream of the shock (c.f. \citealt{beg90}). More directly, the boundary layer X-ray radiation should manifests at the inter-knot sections of the jet. Due to the acceleration taking place continuously within the whole radiating volume, such an X-ray flux should be approximately constant (or change slowly) along the jet. Unfortunately, there are no observations in large scale jets of the counterjet X-ray spectra. In some cases, the lower limits for the X-ray brightness asymmetry can be estimated, like for the 3C 66B ($> 25$, \citealt{hac01}), Pictor A ($> 15$, \citealt{wil01}), or PKS 1127 ($> 5$, \citealt{sie02}). There is evidence of a non-zero X-ray flux from the counterjet in the last object. Note, that in a framework of radiative boundary layer model, the jet-counterjet X-ray brightness asymmetry is smaller than the one in EC models with strong beaming.  

\subsection{Very high energy $\gamma$-ray emission}

Very high energy (VHE) $\gamma$-rays are directly observed only from a few nearby AGNs during their flaring state (c.f. a review by \citealt{kif01}). Very short time scales of the spectral variations at TeV photon energies suggest small sub-parsec sizes of the emitting regions, where the violent particle shock acceleration take place. In our model, the electrons accelerated within the boundary layer of the large scale jets are also able to produce such energetic radiation, mainly due to the IC scattering of the CMB photons, as illustrated on figure 2. One should note, that figure 2 corresponds to the most optimistic scenario, with the highly relativistic jet spine and very efficient acceleration creating the electrons with large Lorentz factors up to $\gamma_{eq} \sim 10^8$. Smaller values of $\Gamma_j$ and $\gamma_{eq}$ would result in decreasing the observed VHE flux and shifting the spectral cut-off to the lower frequencies. Alternatively, for the lower jet velocities or smaller $\gamma_{eq}$ at the few-of-kpc scales the blazar emission illuminating the jet from behind can provide seed photons to produce TeV radiation. Yet another possibility considered in the literature is producing the VHE $\gamma$-rays by the relativistic protons accelerated in the large scale relativistic jets \citep{aha02}.

VHE $\gamma$-rays generated at the distant jets lose their energies due to absorption on the CIB radiation, CMB photons, and other photon fields surrounding the source. \citet{aha94} suggested, that the electromagnetic cascades initiated by such an absorption create the extended {\it pair halo} around the 10-100 TeV extragalactic sources. VHE radiation generated at the jet boundary layer should naturally contribute to this process.

\subsection{Discussion of the free parameters of the model}

In order to illustrate the main physical features of the considered
model we selected its parameters in a way to exhibit clearly its main
characteristics. In particular we neglect possible radiation components
from the jet spine and the ambient medium next to the boundary layer, as well
as from the shocks possibly formed in the flow. In real jets the respective
components add to the observed object radiation.
For the considered radiating electrons we assume energy equipartition
with the background magnetic field and the equal division of energy
between the power-law and the final bump components. One should
note that application of this simple two-component distribution allows for
analytic study of the formed spectrum,  it enables a clear separation of signatures
of these both components in it, but to model the observations a more
elaborated distributions can be required. Such distributions will involve
an initial power-law segment in the spectrum, which flattens at larger energies and forms
a final (not sharp) bump before the exponential cut-off (work in
preparation; cf. \citealp{pet99}). We also take a particular set of
background conditions for the jet and its
boundary layer (the magnetic field and the matter content, the
turbulence and the large scale kinematic structure) and the ambient
radiation fields for this model. Let us consider how the main
predictions of the presented model depend on certain values of the free
parameters assumed in the text, and in particular if the maximum energy
of the boundary layer electrons can substantially differ among different
jet models.

For the efficient particle acceleration within the shear boundary layer
we  assume its turbulent character with high amplitudes MHD or plasma
waves scattering energetic electrons. The assumed configuration of the mean
magnetic field within the shear region with dominating components parallel to the jet axis
was already discussed and seems to be confirmed by the observations. For these
conditions the acceleration process acts as described in Section 2, with
the time scales for particle escape and radiative losses (Eq-s~2.3 and 2.4)
satisfying the inequality $T_{esc} > T_{loss}$, equivalent to $D_{kpc} > 0.04 \,
B_{\mu G}^{-3/2}$. One may note that for a range of  magnetic fields in large scale jets,
$10 < B_{\mu G} < 10^3$, the above condition can be easily satisfied for the transition
layer thickness $D_{kpc} \sim 1$ cited in the literature.

The estimate of the maximum electron energy in Eq.~2.5 gives $\gamma_{eq} \sim 8.7 \cdot
10^7 \, B_{\mu G}^{1/2} \, n_{-4}^{-1/2}$. The parameters in this expression,
the magnetic field and the jet particle density, cannot be varied
arbitrary as they also determine the jet energetics. With the
anticipated model of the relativistic jet dynamically dominated by
cold protons, the jet (spine) kinetic power is given by $L_j \sim \pi \, R_j^2 \, c \,
\Gamma_j^2 \, m_p \, c^2 \, n$ and the comoving magnetic field energy density should
be less than cold protons energy density, $u_B < m_p \,c^2 \, n$. Let us
rewrite the above expressions as $B_{\mu G} < 2 \cdot 10^3 \, n_{-4}^{1/2}$ and $L_{j, 47}
\sim 1.3 \, R_{j, kpc}^2 \, \Gamma_j^2 \, n_{-4}$, where $L_{j, 47} = L_j /
10^{47} \, {\rm erg / s}$. Estimates often give for powerful jets $L_{j, 47} \sim 1$
\citep{sik01}. Hence, the number density $n_{-4} \sim 0.01 - 1$ is estimated for the cases of,
respectively, highly relativistic and nonrelativistic large scale jets with
$R_{j,kpc} \sim 1$ and the maximum magnetic field should be in the range  $10^{-4} -
10^{-3} \, {\rm G}$. In Section 2 we adopted the high value of the relativistic jet density,
$n_{-4} \sim 1$, as the scale of $10^{47} \, {\rm erg / s}$ should be regarded as the lower
limit for the total kinetic energy of the powerful jets (cf. discussion in \citealp{sik01}).
Observations of knots in the large scale jets indicate that their magnetic fields
amplified by the shock compression are $\sim 10^{-5} - 10^{-4} \, {\rm
G}$. The magnetic field within the boundary layer, possibly amplified by the velocity shear,
is likely to be not greater than these values. On the other hand, it is not expected to be
smaller than the magnetic field $\sim 10^{-6} - 10^{-5} \, {\rm G}$ inside
radio lobes. Therefore, we conclude that the most likely value of the magnetic induction
within the boundary layer is close to the assumed above $B_{\mu G} \sim 10$.
However, even if one will adopt the less optimistic scenario of a heavy jet ($n_{-4} \sim
1$) with the weak boundary layer magnetic field ($B_{\mu G} \sim 1$) the
maximum electron energy is still large, $\gamma_{eq} \sim 10^8$, and the
maximum synchrotron frequency $\nu_{syn,eq} \sim B_{\rm \mu G} \, \gamma_{eq}^2 > 10^{16}
\, {\rm Hz}$. For more realistic parameters the synchrotron emission peaks at high X-ray
frequencies.

As a result of weak dependence of $\gamma_{eq}$ on the boundary layer
parameters the expected slowly varying along the jet
radio-to-optical spectral index, the discussed decrease of the jet-counterjet radio
brightness asymmetry and distributed along the jet synchrotron emission
peaking at X-ray frequencies, are expected to hold quite generally.
Presence of the flat high energy electron component
(pile-up bump) also seems to appear in a natural way,
without a particular tuning of the jet parameters. The main
uncertainties of the presented results are connected with the hardly
known velocity profiles of the large scale jets. The bulk Lorentz factor
of the flow, $\Gamma$, determines the relative importance of the
external photon fields cooling electrons via inverse-Compton
scattering. As discussed in Section 3, the energy density of the AGN core
emission exceeds the magnetic field energy density at distances $z_{kpc} <
10 / \Gamma$ from the galactic center, while the CMB energy density is
higher than $u_B$ for $B_{\mu G} < 3 \, \Gamma$ in our local universe.
For the considered values of $1 < \Gamma < 10$ at the tens-of-kpc jet
scales, we expect negligible contribution to the radiative output
from comptonization of the AGN core emission, and a rough equipartition
between the magnetic field and the CMB radiation. It results in
approximately equal fluxes of the synchrotron and the IC boundary layer radiation, if
one neglects possible propagation effects to the observer. The form of the $\Gamma(r)$
radial profile and the related spatial variation of acceleration efficiency can
significantly influence the beaming pattern and intensity of the boundary layer
emission. Therefore the large scale jet spectra at $\gamma$-ray frequencies, in particular
relative normalization of the SSC and EC components presented at figure 2, illustrate
their qualitative features only. The isotropic distribution of the seed photons
assumed in deriving the SSC emission may result in its overestimate. However, even
though comptonization of the CMB photons exceeds the self-Compton emission (for anticipated
$u_B \sim u_{cmb} \sim u_e$), and therefore more detailed treatment of
the SSC radiation, involving anizotropic nature of its seed photons, does not result in
changing of the observed composite high energy spectra.

Finally, let us shortly comment on the jet-counterjet X-ray spectral index
asymmetry and its angular dependence, $\alpha_{X,eff}(\theta^{\ast})$ (Eq.~4.2).
Although the hard electron component is always expected to dominate the boundary layer X-ray
emission, the observed spectral index at critical frequencies can substantially vary for
different jet viewing angles, the jet bulk Lorentz factors and the exact shape of the
electron spectrum at highest energies. Therefore, figure 4 illustrates only a possible
behavior of the X-ray jets demonstrating different spectral indices of the jet and the
counterjet.

\section{Conclusions}

In the present paper, we study radiation of the ultrarelativistic electrons accelerated at  boundary shear layers of large scale relativistic jets. The acceleration process acting at such layer is substantially different from the often considered first order shock acceleration. It can generate, in a natural way, a characteristic two component particle spectrum with a low energy power-law distribution extending in energy up to the harder pile-up component forming at higher energies due to radiation losses. In order to study a possible role of the mentioned process we constructed a simple, in its several aspects qualitative model of particle acceleration and radiation from the jet boundary. We propose, that the emission of such electrons can substantially contribute to the jet radiative output and we investigate the resulting observational effects. Modifications of the standard jet model arise from the kinematic effects involved (c.f. \citealt{kom90,lai99,chi00,cel01}) and the mentioned nature of the acceleration process \citep{ost00}:
\begin{itemize}
\item With the considered above continuous {\it in situ} acceleration model, the radio-to-optical range of the boundary layer emission is characterized by weak variation of the spectral index along the jet, assuming the boundary layer physical parameters change smoothly with the distance from the nucleus. The considered radial velocity profile of the relativistic jet defines the beaming pattern of the boundary layer emission. In general, the velocity shear within the boundary region results in decreasing the observed jet-counterjet brightness asymmetry, affecting evaluations of the jet bulk Lorentz factors.
\item If the considered cosmic ray population involves the high energy bump electrons with the energy density comparable to the magnetic field energy density, one should observe a high frequency (X-ray) excess over the power-law radiation extrapolated from the lower frequency measurements. In a framework of our model, the spectral properties of the X-ray bump emission (the effective spectral index, jet-counterjet brightness asymmetry, etc) can substantially differ from the one observed at lower frequencies. We speculate, that such features can be seen at least in some Chandra jets.
\item The considered model can explain the observed in some cases edge-brightened jets in a natural way.
\end{itemize}
One should note, that our conclusions depend weakly on the unknown detailed physical conditions within the transition layer. The model presented here, far from being complete, is intended to describe the main radiative properties of the boundary region under simple assumptions about turbulent character of the shear layer, relativistic jet velocities on the large scales and energy equipartition between the magnetic field and the accelerated electrons. We expect, that careful analysis of observational data should reveal the effects
discussed above and thus provide restrictions both on physical parameters
within such boundary layers and the proposed model of electron acceleration.

\acknowledgments

We are grateful to Marek Sikora for his help and discussions. MO acknowledges a useful discussions with Vah\'{e} Petrosian. Remarks of the anonymous referee enabled to substantially improve the final version of the paper. The present work was supported by Komitet Bada\'{n} Naukowych through the grant BP 258/P03/99/17.

\clearpage

\begin{figure}
\plotone{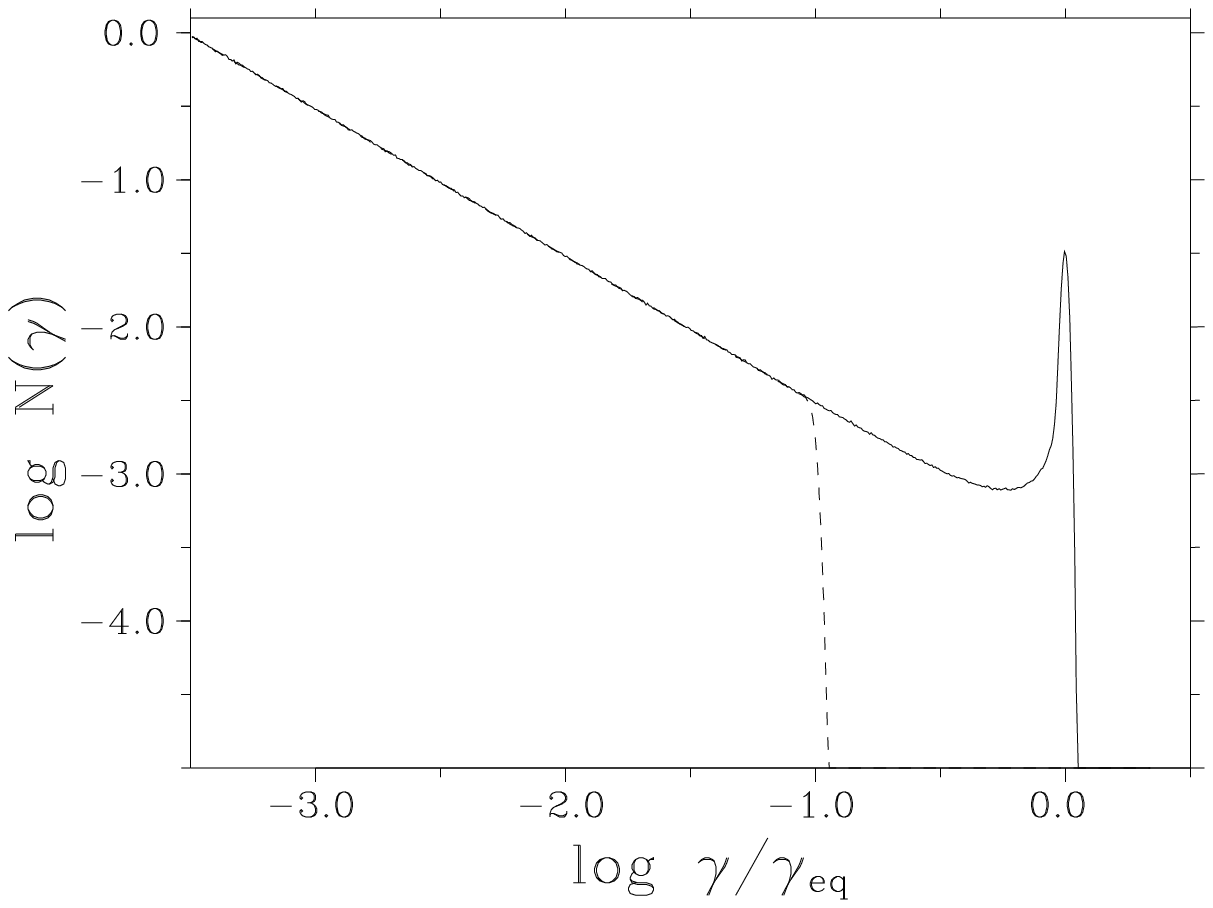}
\caption{ A simple model spectrum $N(\gamma) \equiv d \ n_{e} (\gamma) / d \ log \ \gamma$ of particles accelerated within the turbulent shear layer. The acceleration process creates power-law energy distribution ended by the cut-off (the dashed line) or by the high energy pile-up bump at the maximum electron energy $\gamma_{eq}$ (the soild line) (cf. \citealt{ost00}). \label{fig1}}
\end{figure}

\clearpage

\begin{figure}
\plotone{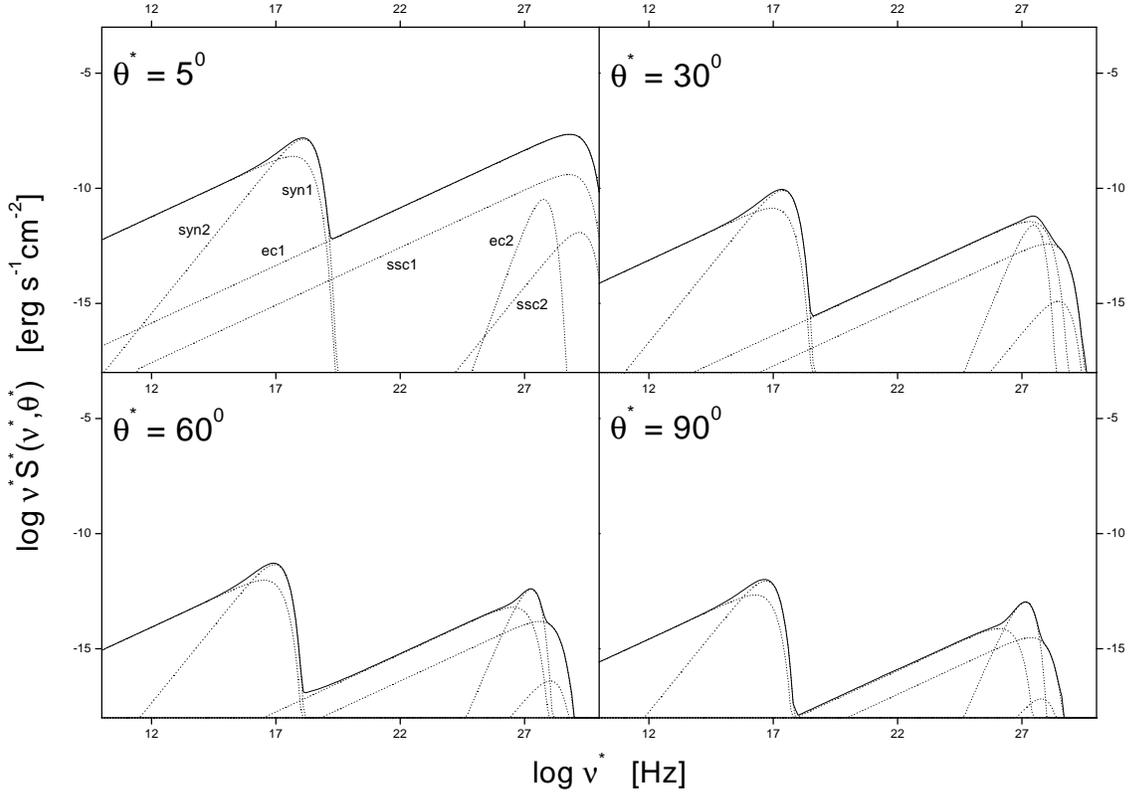}
\caption{ The observed spectral energy distributions of the radiation generated within the boundary shear layer for $\gamma_{eq} = 10^8$ and $\theta^{\ast} = 5^0 \, , \, 30^0 \, , \, 60^0 \, , \, 90^0$, at the respective panels. The presented spectra are integrated over the flow Lorentz factor profile (3.9) and correspond to the parameters discussed in the text. The index `1' refers to the emission $\propto \nu^{-1/2}$ of the power-law part of the electron spectrum. The index `2' refer to the emission of the high energy electron bump component. In the case of the SSC process, monoenergetic electrons contribute also to the emission $\propto \nu^{-1/2}$. Absorption of VHE $\gamma$-rays during the propagation to the observer is neglected. \label{fig2}}
\end{figure}

\clearpage

\begin{figure}
\plotone{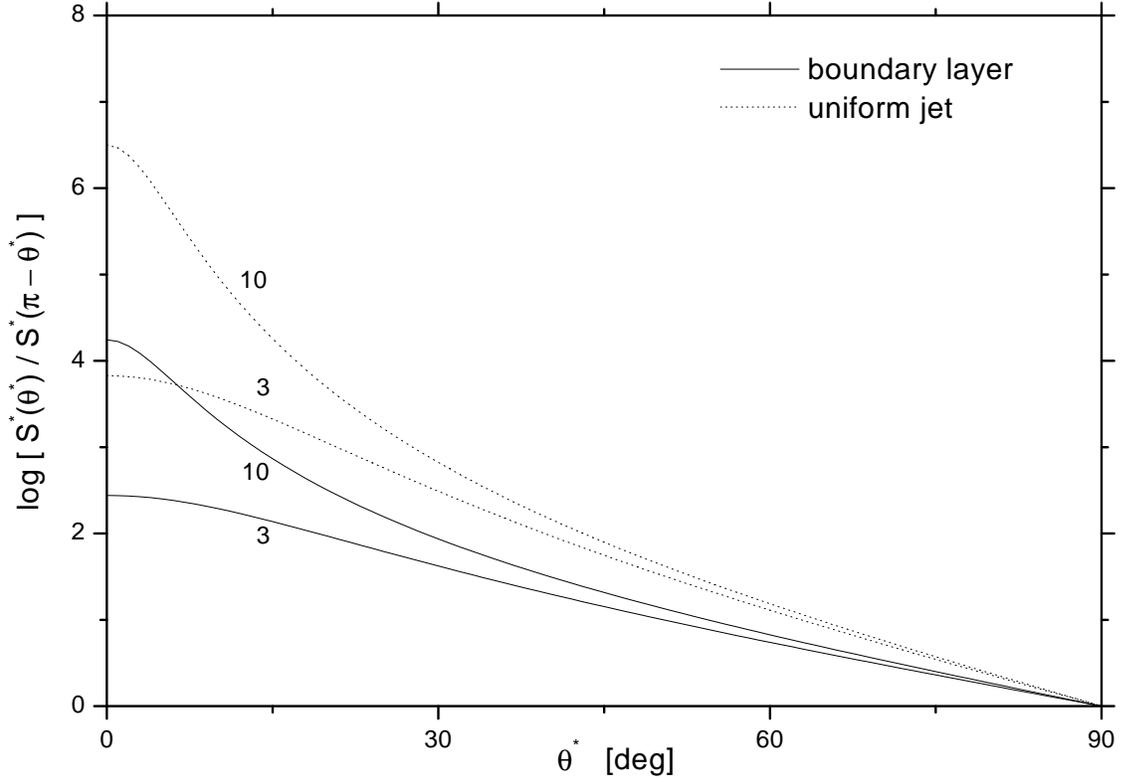}
\caption{ A ratio of the observed radio fluxes from the jet and the counterjet, $S^{\ast}(\theta^{\ast}) / S^{\ast}(\pi - \theta^{\ast})$, as a function of the viewing angle $\theta^{\ast}$. The dotted lines correspond to the model with the uniform flow Lorentz factor $\Gamma_j = 10$ or $3$ within the whole jet. The solid lines correspond to the radiation from a boundary shear layer with the radial profile (3.9) and the same $\Gamma_j = 10$ or $3$. The Lorentz factors are provided near the respective curves. \label{fig3}}
\end{figure}

\clearpage

\begin{figure}
\plotone{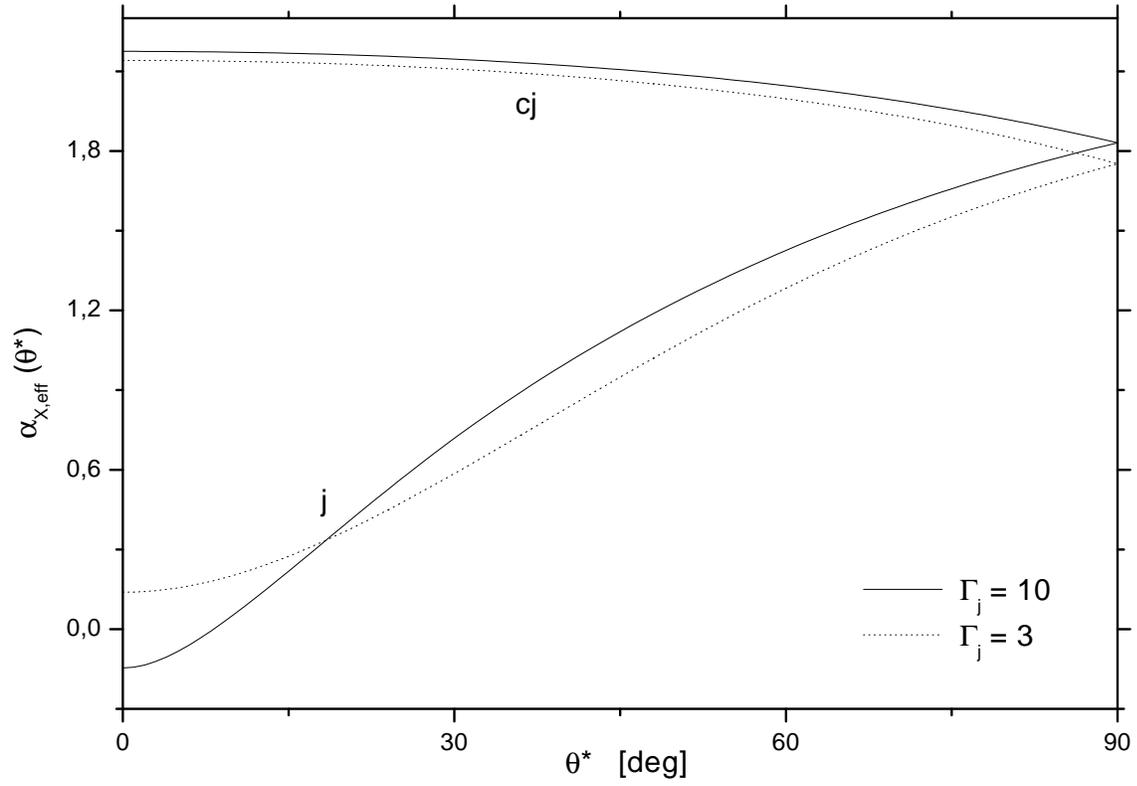}
\caption{ An illustration of the possible effective X-ray spectral index of a jet (`$j$') and a counterjet (`$cj$') boundary layer emission, $\alpha_{X, eff}$, as a function of the viewing angle $\theta^{\ast}$. The solid lines correspond to $\Gamma_j = 10$ and the dashed ones to $\Gamma_j = 3$. \label{fig4}}
\end{figure}

\end{document}